\documentclass[graybox]{svmult}

\usepackage{mathptmx}       % selects Times Roman as basic font
\usepackage{helvet}         % selects Helvetica as sans-serif font
\usepackage{courier}        % selects Courier as typewriter font
\usepackage{type1cm}        % activate if the above 3 fonts are
\usepackage{url}
\usepackage{makeidx}         % allows index generation
\usepackage{graphicx}        % standard LaTeX graphics tool
\usepackage{multicol}        % used for the two-column index
\usepackage[bottom]{footmisc}% places footnotes at page bottom

\makeindex             % used for the subject index
\begin{document}

\title*{Quality Classifiers for Open Source Software Repositories}
\author{George Tsatsaronis and Maria Halkidi and Emmanouel A. Giakoumakis}
\institute{George Tsatsaronis \at Department of Informatics,
Athens University of Economics and Business, 76, Patision Str.,
Athens, \email{gbt@aueb.gr} \and Maria Halkidi \at Department of
Technology Education and Digital Systems, University of Pireaus,
\email{mhalk@unipi.gr} \and Emmanouel A. Giakoumakis \at
Department of Informatics, Athens University of Economics and
Business, 76, Patision Str., Athens, \email{mgia@aueb.gr} }
\maketitle
\vspace{-0.2in}
\abstract{Open Source Software (OSS) often relies on large
repositories, like SourceForge, for initial incubation. The OSS
repositories offer a large variety of meta-data providing
interesting information about projects and their success. In this
paper we propose a data mining approach for training classifiers
on the OSS meta-data provided by such data repositories. The
classifiers learn to predict the successful continuation of an OSS
project. The `successfulness' of projects is defined in terms of
the classifier confidence with which it predicts that they could
be ported in popular OSS projects (such as FreeBSD, Gentoo
Portage).}

%The classifiers can assist with predicting the future of any
%submitted OSS project(i.e. whether the project will be ported by
%other popular OSS projects). We argue that this new aspect of
%measuring successfulness of OSS projects can be added as an
%additional metric in previously proposed models of OSS
%successfulness. We have experimentally evaluated the proposed
%approach in the SourceForge and the FreshMeat project data
%collected by the FLOSS project. The reported results are promising
%and demonstrate the significance of the information that OSS
%repository meta-data can provide.

%-----------------------------------------------------------------
%-----------------SECTION 1: Introduction-------------------------
%-----------------------------------------------------------------
\vspace{-0.2in}
\section{Introduction}
\vspace{-0.1in}
Initial open source software (OSS)
projects rely on large repositories for hosting and distribution
until they become independent. A huge amount of project meta-data
is collected and maintained in such software repositories
providing useful information about projects and their success. In
this paper we propose a data mining approach that processes the
meta-data contained in such OSS repositories. The proposed
approach aims at the construction of a classifier that is trained
on the meta-data of existing projects and predicts the successful
continuation of any given OSS. The \textit{successfulness} of a
project is defined with regard to the confidence level of the
classifier which predicts that this project will be ported in
widely used OSS projects (e.g. FreeBSD). We argue that the
classifier decision, along with its confidence level, can be
incorporated into known models of software success, like the model
of DeLone and McLean \cite{DeLone:92}, or its reexamination and
expansion for OSS software, by Crowston et al. \cite{Crowston:06}.
We have experimentally evaluated the proposed mining approach in
the SourceForge and the FreshMeat OSS project meta-data released
from the Floss project. We also evaluated the importance of the
underlying features using Information Gain and Chi-Square. The
results of this study report high F-Measures for the classifiers
based on the most important FLOSS features.

The proposed approach consists of two main steps: a) data
collection and pre-processing, b) training classifiers. The
meta-data in OSS repositories are usually provided in formats that
are not suitable for mining. Thus, one of the most important
elements in the proposed approach is the \textit{pre-processing
procedure}. Techniques such as \textit{parsing, crawling} and
\textit{feature selection} are used to collect data from the FLOSS
project, which contains crawled projects from important OSS
repositories, like SourceForge and FreshMeat. We also discuss
methods that can be used to train classifiers on the stored project data.
%with which classifiers can train and be applied on the stored project data.

The rest of the paper is organized as follows. Section
\ref{sec:related} discusses related work in mining OSS projects
and related models for measuring information systems' success.
Section \ref{sec:framework} presents the data mining approach for
extracting knowledge from OSS repositories. Section
\ref{sec:experiments} provides experimental results and analysis.
Section \ref{sec:conclusions} concludes and gives further insight
to possible future work. \vspace{-0.2in}
\section{Related Work}\label{sec:related}
\vspace{-0.05in} \textbf{Data Mining in Software Engineering}.
Data mining is widely used for supporting industrial scale
software maintenance, debugging and testing. An approach that
exploits classification methods to analyze logical bugs is
proposed in \cite{LY+06}. This work treats program executions as
software behavior graphs and develops a method to integrate closed
graph mining and SVM classification in order to isolate suspicious
regions of non-crashing bugs. A semi-automated strategy for
classifying software failures is presented in \cite{Pod03}. This
approach is based on the idea that if $m$ failures are observed
over some period during which the software is executed, it is
likely that these failures are due to a substantially smaller
number of distinct defects. A predictive model for software
maintenance using data and text mining techniques is proposed in
\cite{RT07}. To construct the model, they use data collected from
more than $100.000$ open source software projects lying in the
SourceForge portal. Using SAS Enterprise Miner and SAS Text Miner,
they focused on collecting values for variables concerning
maintenance costs and effort from OSS projects, like Mean Time to
Recover (MTTR) an error. They clustered the remaining projects
based on their descriptions, in order to discover the most
important categories of OSS projects lying in the SourceForge
database. Finally, they used the SAS Enterprise Miner to train
classifiers on the MTTR class variable. The reported results
highlight interesting correlations between the class variable and
the number of downloads, the use of mail messages and the project
age. There is also a number of other works \cite{Williams05,
Bowring04, Francis04, Pod03, LY+06} that use
classification methods in software engineering, assisting with its
main tasks (development, debugging, maintenance, testing).\\

\noindent \textbf{Models of OSS Success}.\label{sec:models} The
most popular model for measuring information systems' (IS) success
is the one proposed by DeLone and McLean \cite{DeLone:92}. They
actually introduce six interrelated factors of success: 1) system
quality, 2) information quality, 3) use, 4) user satisfaction, 5)
individual impact, and 6) organizational impact. Based on this
approach, Seddon \cite{Seddon:97} reexamined the factors that can
measure success and concluded that the related factors are system
quality, information quality, perceived usefulness, user
satisfaction and IS use. Based on those approaches, a number of
measures that can be used to assess the success in FLOSS is
presented by Crowston et al. in \cite{Crowston:06}. These measures
are defined based on the results of a statistical analysis applied
to a subject of project data in FLOSS. Specifically the empirical
study was based on a subset of SourceForge projects. In this paper
we propose another such measure, that can be added to the
\emph{use} and the \emph{user satisfaction} factors of the
proposed models of success. %To the best of our knowledge, it is
%the first approach that uses classification techniques to assess
%the quality of software projects.
\vspace{-0.2in}
\section{Software Success Classification Approach}
\vspace{-0.05in}
\label{sec:framework} We introduce a classification approach that
adopts data mining techniques in order to extract useful
information from OSS repositories and further analyze it to
predict softwares' successful continuation. Based on that, our
approach aims at constructing a metric that assesses software
success and which can be considered as an implementation of the
factors \emph{use} and \emph{user satisfaction} to the model of
DeLone and McLean \cite{DeLone:92} for measuring IS success. An
expansion of this model has been proposed by Crowston et al.
\cite{Crowston:06} for OSS software, according to which the data
lying in FLOSS from SourceFourge are mapped to potential measures
of OSS success. The following metric can be incorporated to the
\emph{System use} process phase, as this is described in
\cite{Crowston:06}. We have developed the proposed metric taking
into account the FLOSS data for both SourceForge and FreshMeat
repositories.
\vspace{-0.2in}
%http://www.freebsd.org/ports/master-index.html
\subsection{OSS Porting Classification Metric}
\vspace{-0.1in} Without loss of generality we consider that OSS
software ported to widely used open source operating systems is
widely accepted as useful and significant by the majority of
users. Then we claim that a project is considered to be
\textit{successful}, from the point of view of popularity and user
satisfaction, if it is selected to be ported in two of the most
popular open source operating systems, namely FreeBSD and Gentoo.
Based on this, we construct classifiers, and we use the confidence
of their classification output as a metric of OSS successfulness.
The main modules of our approach can be summarized as follows: (a)
\textit{Data collection and pre-processing}. The OSS repositories maintain
huge amount of OSS meta-data that provide useful information about
the hosted projects. This module refers to methods used to collect
data from OSS repositories and properly pre-process them for data
mining.(b) \textit{Software classification}. This module provides the
methods to train classifiers for predicting projects' course over
time, trained on the collected OSS meta-data. The classifiers are
built to predict the successful continuation of an OSS based on a
specific set of project features. Based on the trained
classifiers, new project releases (or unclassified projects) can
be classified with regard to their successfulness (as this has
been defined above).\\

%\subsubsection{Data Collection and Preprocessing Techniques}
\noindent
\textbf{Data Collection and Preprocessing Techniques}\\
A large amount of data are collected and maintained in OSS
repositories. These data contain useful information about projects
that we aim to analyze so as to extract interesting knowledge and
make inferences about future course of projects. However the data
are provided in various formats that in most of the cases are not
suitable for data mining. Thus, a pre-processing procedure is
needed before data mining is applied to the available data.
Specifically, techniques such as \textit{crawling} and {\it
feature selection} are used to collect project data from various
OSS portals and select those that contain interesting information
for further analysis. \textit{Crawling} usually refers to the
process of browsing the World Wide Web in a methodological and
automated manner. An extensive analysis of a crawling mechanism is
provided by Chakrabarti in Chapter 2 of \cite{Chakrabarti:2003}.
We used a smaller crawling mechanism to browse the Web pages of
the two open source operating systems (FreeBSD, Gentoo) and
collect further information about the projects existing in
SourceForge and FreshMeat for later processing. Many of the
programming techniques used are described in \cite{Loton:2002}.
For the processing of the SourceForge and the FreshMeat data, we
used the FLOSS data and index. \textit{Feature selection} is the
technique of selecting a subset of relevant features for creating
robust learning models. In our approach, feature selection
techniques assist in a two-fold manner. Firstly, they assist the
mining procedure with further analyzing the project data crawled.
Thus, they provide an image of all the features being used.
Secondly they assist in selecting the features that are more
useful with regards to the final classification, which in our case
is to predict whether an OSS software is successful enough to be
ported in FreeBSD and Gentoo Portage. As feature selection
criteria, we have adopted Information Gain (IG) and Chi-Squared
($x^2$).

According to the IG criterion, the expected reduction of
uncertainty in guessing the class variable, once the feature value
is known, needs to be measured. This is expressed through
measuring the expected reduction in entropy, once the feature
value is known, given by equation ~\ref{eq:IG}. While for the case
of discrete value features equations 1-2 apply, in the case of the
continuous value features, the domain of each feature variable $X$
is divided into many subintervals of a given equal length and each
$X_i$ is true iff X belongs to the corresponding interval. Let $S$
be the set of s data objects. Considering a set of $m$ classes
$C_i$, the expected information needed to classify a given object
is defined as follows:

\vspace{-0.15in}
\begin{small}
\begin{equation}
    I(s_1, \ldots, s_m) = - \sum_{i=1}^{m} p_i log_2(p_i)
\end{equation}
\end{small}
\vspace{-0.05in}
where $s_i$ is the number of data objects in S labeled $c_i$ and
$p_i$ is the probability that an object belongs to class $c_i$, $
p_i = s_i/s$. Let attribute A have $v$ distinct values
$\{\alpha_1, \alpha_2, \ldots, \alpha_v\}$. The attribute $A$ can
be used to partition $S$ in $v$ subsets $\{S_i\}^{v}_{i=1}$, where
$S_i$ contains those objects in S that have value $\alpha_i$ for
$A$. The entropy or expected information based on the partitioning
of $S$ into subsets by $A$ is given by:

\vspace{-0.1in}
\begin{small}
\begin{equation}
    E(A) = \sum^{v}_{j=1} \frac{s_{1j}+\ldots+s_{mj}}{s}-\sum_{i=1}^{m} p_{ij}log_2(p_{ij}),
    \ldots, s_{mj})
\end{equation}
\end{small}
\vspace{-0.05in}
where $s_{ij}$ is the number of data objects of class $C_i$ in a subset $S_j$, and
%\vspace{-0.1in}
%\begin{equation}
%I(s_{1j}, \ldots, s_{mj}) = -\sum_{i=1}^{m} p_{ij}log_2(p_{ij})
%\end{equation}
%\vspace{-0.05in}
%where
$p_{ij}=\frac{s_{ij}}{\left|S_j\right|}$ is the
probability that an object in $S_j$ belongs to class $C_i$. The
encoding information that would be gained by branching on A is:

\vspace{-0.1in}
\begin{small}
\begin{equation}
    Gain(A) = I(s_{1j}, \ldots, s_{mj}) - E(A) \label{eq:IG}
\end{equation}
\end{small}
\vspace{-0.1in}

Chi-squared ($x^2$) is often used to measure the association
between two features in a contingency table. In the case of a
binary classification problem it can be used to measure the
association between an input feature and the class variable, as
shown in the following equation:

\vspace{-0.1in}
\begin{small}
\begin{equation}
x^2=\sum_{i=1}^{r}{(\frac{(n_{iP}-
\mu_{iP})^2}{\mu_{iP}}+\frac{(n_{iN}- \mu_{iN})^2}{\mu_{iN}}) }
\end{equation}
\end{small}
where $r$ are all the possible values of the examined feature, $N$
and $P$ are the two classes (Negative and Positive respectively),
$n_{iP}$ and $n_{iN}$ are the number of instances belonging to
class $P$ or $N$ respectively and have the value $i$ of the
examined feature, and $\mu_{ij}=\frac{n_{*j}n_{i*}}{n}$ with n
being the number of instances, $i=1,\ldots,r$ and $j=P,N$. In this
work we use both IG and chi-square to measure the significance of
the stored features contained in the OSS projects' meta-data, as
crawled by the FLOSS project.\\

\noindent
%\subsubsection{Software Classification}
\textbf{Software Classification}\\
%\label{sec:classifiers}
In order to train classifiers that can predict the degree of
successfulness of an OSS lying in FreshMeat or SourceForge, the
most important factor is to define the projects that belong to the
classes of \textsl{successful} and \textsl{unsuccessful} projects.
This can be the training set of projects that can be used for
training the classifiers. The approach we adopt is to define as
\emph{successful} the OSS projects that have been ported in both
FreeBSD and Gentoo Portage. This heuristic criterion satisfies
part of the \emph{System use} process phase of the model of
successfulness defined in \cite{Crowston:06}, namely
\emph{Interest}, \emph{Number of Users} and \emph{User
Satisfaction}. We concluded to that criterion after having crawled
the FreeBSD, Gentoo Portage and the Debian Popularity Contest,
aiming to find their common set of projects with the considered
FLOSS projects. FreeBSD
Ports\footnote{\url{http://www.freebsd.org/}} is a package
management system for the FreeBSD operating system and it contains
all projects that are ported to FreeBSD.
Portage\footnote{\url{http://www.gentoo.org/}} is also a package
management system but it contains projects ported to Gentoo Linux.
Furthermore, Debian Popularity
Contest\footnote{\url{http://popcon.debian.org/}} is a project
which attempts to map the usage of Debian packages. For the Gentoo
Portage, we have used the latest
snapshot\footnote{\url{http://gd.tuwien.ac.at/opsys/linux/gentoo/snapshots/}}
in order to determine the projects that are ported into it. From
the Debian Popularity Contest we have collected for each package
of Debian the number of users who installed it, the number of
users who use it frequently, the number of users who do not use it
frequently, the number of users who upgraded to the latest version
of the software and the number of users that did not publish
enough information. After having carefully analyzed all these data
and examined several criteria that can be used as determining
\emph{successful} and \emph{unsuccessful} projects, we concluded
that the aforementioned criterion (porting of an OSS project in
both FreeBSD and Gentoo Portage) expressed better the \emph{System
Use} as discussed earlier, and also provided better experimental
results that other criteria examined.

Having determined the \emph{successfulness} criterion, it is
straightforward to construct classifiers taking into account the
features offered by SourceForge and FreshMeat (an analysis of the
features follows in the next section). Classification is a two
step process:
\begin{enumerate}
    \item \textit{Training step. }A classification model is built describing a predefined set of classes
    (i.e \emph{Successful}, \emph{UnSuccessful}).
     The model is trained by analyzing a set of data whose classification (class labels) is known (training data set).
  \item \textit{Classification step}. First the predictive accuracy of the trained model (classifier) is estimated.
  If it is acceptable the classifier is used to classify project instances for which the class label is unknown.
\end{enumerate}

Since each repository (FreshMeat, SourceForge) maintains different
attributes for the hosted projects, a classifier per repository
must be developed. Based on our approach any of the widely used
classification algorithms can be used to analyze the training set
of projects and construct the software classification model. In
our current work, we have adopted \textit{Naive Bayesian, Decision
trees} and \textit{Support Vector Machines} classification methods
to train three different classifiers based on the available set of
project data. For each of these classifiers, the successfulness
metric is their confidence level of the classified instance. For
the SVM, it is the distance of the considered project feature
vector, from the hyperplane separating negative from positive
instances. For the NB (Naive Bayesian) classifier, it is the
probability that the considered project feature vector belongs to
the \emph{successful} class. Finally, for the decision trees (i.e.
the C4.5 classifier) it is the frequency of occurrence of the
training examples that are in the \emph{successful} class,
following the branch of the tree with attribute values of the
examined instance. The experimental study in section \ref{sec:experiments} shows the
performance of all three used classifiers for both FreshMeat and
SourceForge.
\vspace{-0.25in}
\subsection{Features Description}
\label{sec:fdescription} \vspace{-0.1in} In this work we have used
project data that have been collected and are available from the
FLOSSMole project
\footnote{\url{http://flossmole.sourceforge.net/}}. Specifically
we work with the releases of SourceForge and FreshMeat project
data indexed by the FLOSS repository. Below we present the main
attributes (features) of the project data in the FLOSS repository
that are used for the classification process. Also we indicate
which portal (FreshMeat - FM or SourceForge - SF) maintains each
attribute for the projects it hosts.

\begin{enumerate}
    \item \textit{Project License (FM and SF):} Project's license type (such as GPL, LGP, BSD License, Freeware, Shareware etc)
    \item \textit{Vitality Score} (FM) which is defined as
   % \begin{equation}
   $ Vitality Score = \frac{V * T_0}{T_n}$\\
    %\end{equation}
    where $V$ is the number of the project's versions, $T_0$ is
    the time elapsed since first project upload (usually counted in
    days), and $T_n$ is the time elapsed since latest version
    upload.
    \item \textit{Popularity Score (FM): }Represents project's popularity based on:
    \begin{itemize}
        \item Users' visits in project's URL, i.e. URL hits(further referred to as $a$).
        They refer to the visits in project's original web site, and not at FreshMeat's site for the project.
        \item Users' visits in FreshMeat's project site ($b$).
        \item Users' subscriptions ($c$).
    \end{itemize}
    Then, the popularity score is measured as: $ Popularity Score = \sqrt{(a+b)\cdot(c+1)}$

    \item \textit{Rating (FM):} Any subscribed user can rate a project. Given 20 or more user
    ratings, the project engages a ratings based ranking. Rating is measured as a means
    of a weighted ranking (WR), which is computed as follows:
    \vspace{-0.1in}
    \begin{equation}
     WR=(v ÷ (v+m)) × R + (m ÷ (v+m))\cdot C
    \end{equation}
     \begin{small}
     where:
      $R$ = average (mean) rating for the project from users = (Rating)\\
      $v$ = number of votes for the project = (votes)\\
      $m$ = minimum votes required (currently 20)\\
      $C$ = the mean vote across the whole report\\
    \end{small}
    \item \textit{Subscriptions (FM):} Number of subscribed users in a project.
    \item \textit{Developers} (FM and SF): Number of developers per project and other information about them.
    \item \textit{Target audience / End Users} (SF):  The target group of the resulting software.
    \item \textit{Executing operating system} (SF): The operating system in which the resulting project software can execute.
    \item \textit{DBMS Environment/Technology} (SF): For the projects using a DBMS, this is the used DBMS environment, or technology in general.
    \item \textit{Programming Language} (SF): The programming language in which the project software is written.
    \item \textit{Number of downloads} (SF).
    \item \textit{Interface }(SF): The interface of the project (Library, Command Line, Web, GUI etc).
    \item \textit{Natural language }(SF): The natural language of the project (English, France, etc).
    \item \textit{Topic }(SF): The topic of the project (Databases, Network Administration).
    \item \textit{Registration Date} (FM and SF): The date that the project was inserted into the portal.
    \item \textit{Days since Registration Date} (FM and SF): Days elapsed since the registration date.
    \item \textit{Project status} (SF): The status of the project (such as Beta, Planning, Production/Stable, Alpha, Pre-Alpha etc)
    \item \textit{Number of project donators} (SF).
    \item \textit{Rank} (SF): A ranking of the projects produced
    by SourceForge, considering downloads and days since
    registration date.
\end{enumerate}
\vspace{-0.2in}
\begin{table}[b]
\begin{center}
\begin{tabular}{|c||c|c|}
\hline
&\multicolumn{2}{c|}{FreshMeat}\\
\cline{2-3}
&\footnotesize{Information Gain}&\footnotesize{Chi-Square}\\
\hline \hline
\footnotesize{Popularity}&\footnotesize{\textbf{0.324}}&\footnotesize{\textbf{1793.254}}\\
\hline
\footnotesize{Subscriptions}&\footnotesize{\textbf{0.319}}&\footnotesize{\textbf{1756.487}}\\
\hline
\footnotesize{Vitality}&\footnotesize{\textbf{0.238}}&\footnotesize{\textbf{1781.661}}\\
\hline
\footnotesize{\#Rating}&\footnotesize{\textbf{0.219}}&\footnotesize{\textbf{1253.699}}\\
\hline
\footnotesize{Rating}&\footnotesize{\textbf{0.189}}&\footnotesize{\textbf{111.144}}\\
\hline
\footnotesize{Days}&\footnotesize{\textbf{0.107}}&\footnotesize{\textbf{620.316}}\\
\hline
\footnotesize{Developers}&\footnotesize{\textbf{0.05}}&\footnotesize{\textbf{269.701}}\\
\hline
\footnotesize{License}&\footnotesize{\textbf{0.033}}&\footnotesize{\textbf{187.66}}\\
\hline
\end{tabular}
\begin{tabular}{|c||c|c|}
\hline
&\multicolumn{2}{c|}{SourceForge}\\
\cline{2-3}
&\footnotesize{Information Gain}&\footnotesize{Chi-Square}\\
\hline \hline
\footnotesize{Downloads}&\footnotesize{\textbf{0.199}}&\footnotesize{\textbf{759.95}}\\
\hline
\footnotesize{Rank}&\footnotesize{\textbf{0.153}}&\footnotesize{\textbf{595.337}}\\
\hline
\footnotesize{OS}&\footnotesize{\textbf{0.145}}&\footnotesize{\textbf{558.826}}\\
\hline
\footnotesize{Language}&\footnotesize{\textbf{0.138}}&\footnotesize{\textbf{533.17}}\\
\hline
\footnotesize{Days}&\footnotesize{\textbf{0.119}}&\footnotesize{\textbf{469.172}}\\
\hline
\footnotesize{Status}&\footnotesize{\textbf{0.095}}&\footnotesize{\textbf{381.716}}\\
\hline
\footnotesize{Interface}&\footnotesize{\textbf{0.078}}&\footnotesize{\textbf{317.054}}\\
\hline
\footnotesize{Developers}&\footnotesize{\textbf{0.075}}&\footnotesize{\textbf{298.099}}\\
\hline
\footnotesize{Users}&\footnotesize{\textbf{0.072}}&\footnotesize{\textbf{276.279}}\\
\hline
\footnotesize{License}&\footnotesize{\textbf{0.03}}&\footnotesize{\textbf{112.715}}\\
\hline
\footnotesize{DBMS}&\footnotesize{\textbf{0.01}}&\footnotesize{\textbf{5.548}}\\
\hline
\footnotesize{Donors}&\footnotesize{\textbf{0}}&\footnotesize{\textbf{0}}\\
\hline
\end{tabular}
\caption{Information Gain and Chi-Square for the FreshMeat and
SourceForge Features.}
\end{center}
\end{table}
\section{Experimental Evaluation} \label{sec:experiments}
\vspace{-0.1in}

\textbf{Experimental Setup}. The projects that we
considered for our experiments numbered $112915$ for SourceForge
and $41908$ for FreshMeat. For both portals, we found the projects
which are available from FreeBSD Ports and Gentoo Portage in order
to label them as {\it successful}. For our experiments, we used
10-fold cross validation on three classifiers (decision trees,
Naive Bayes, and SVM). For learning decision trees, we used the
J48 algorithm that is WEKA's \cite{Witten:05} implementation of
the C4.5 (an extension of the basic ID3 algorithm) decision tree
learner. We also used WEKA for the Naive Bayesian classifier. For
support vector machines (SVMs) we used Joachim's
$SVM^{light}$\cite{Joachims:99}. For each of the two data sets
(FreshMeat and SourceForge), we trained all three classifiers on
all of the features described. For the features evaluation IG and
chi-square was used. For the evaluation of the classifiers, we
measured precision, recall and F-Measure. The goal of this
experiment is to prove the value of the OSS portals' meta-data,
and consequently the value of the proposed success metric that is
based on training classifiers, as an additional
factor of OSS success.\\

\noindent \textbf{Features Analysis}. Feature selection requires
analysis of all considered features. We have computed IG and
chi-square values of all features using 10-fold cross validation,
based on the used criterion. The measurements for the IG and the
chi-square criteria are shown in Table 1 for the FreshMeat  and
the SourceForge repositories. The results are shown in decreasing
order of importance based on the IG measure. From the results
obtained we note primarily that the produced feature rankings
based on the two measures (IG and chi-square) are exactly the same
in the case of SourceForge, while in the case of FreshMeat the
only discrepancy is that the third feature according to IG is
ranked second according to chi-square. This shows that the
selected criterion of successfulness (porting of an OSS to FreeBSD
and Portage) produces a stable ranking of features for both IG and
chi-square. Regarding the features' importance, in the case of the
FreshMeat data, the top 5 features proved to be {\it popularity},
{\it subscriptions}, {\it vitality}, {\it number of ratings} and
{\it ratings}, while in the case of the SourceForge data, these
are {\it downloads}, {\it rank}, {\it OS}, {\it language} and {\it
days}. The IG drops dramatically for the rest features. At this
point we must note that the top ranked features according to both
measures include the features we were expecting to rank high,
based on previously proposed models \cite{Crowston:06}. The
feature ranking can be used to decrease the classifiers' model
size. In the next section we show that the learned models can be
reduced to only considering the top 5 features for each
repository, without important decrease in performance.\\
\begin{table}[b]
\begin{center}
\begin{tabular}{|c||c|c|c||c|c|c|}
\hline
&\multicolumn{3}{c||}{FreshMeat}&\multicolumn{3}{c|}{SourceForge}\\
\cline{2-4} \cline{5-7}
&\footnotesize{Precision}&\footnotesize{Recall}&\footnotesize{F-Measure}&\footnotesize{Precision}&\footnotesize{Recall}&\footnotesize{F-Measure}\\
\hline \hline
\footnotesize{SVM All Features}&\footnotesize{\textbf{0.88}}&\footnotesize{\textbf{0.57}}&\footnotesize{\textbf{0.69}}&\footnotesize{\textbf{0.67}}&\footnotesize{\textbf{0.75}}&\footnotesize{\textbf{0.7}}\\
\hline
\footnotesize{SVM Top-5 Features}&\footnotesize{\textbf{0.62}}&\footnotesize{\textbf{0.96}}&\footnotesize{\textbf{0.75}}&\footnotesize{\textbf{0.66}}&\footnotesize{\textbf{0.73}}&\footnotesize{\textbf{0.69}}\\
\hline
\footnotesize{C4.5 All Features}&\footnotesize{\textbf{0.79}}&\footnotesize{\textbf{0.81}}&\footnotesize{\textbf{0.79}}&\footnotesize{\textbf{0.77}}&\footnotesize{\textbf{0.71}}&\footnotesize{\textbf{0.73}}\\
\hline
\footnotesize{C4.5 Top-5 Features}&\footnotesize{\textbf{0.77}}&\footnotesize{\textbf{0.78}}&\footnotesize{\textbf{0.77}}&\footnotesize{\textbf{0.75}}&\footnotesize{\textbf{0.7}}&\footnotesize{\textbf{0.72}}\\
\hline
\footnotesize{NB All Features}&\footnotesize{\textbf{0.76}}&\footnotesize{\textbf{0.83}}&\footnotesize{\textbf{0.79}}&\footnotesize{\textbf{0.81}}&\footnotesize{\textbf{0.78}}&\footnotesize{\textbf{0.79}}\\
\hline
\footnotesize{NB Top-5 Features}&\footnotesize{\textbf{0.74}}&\footnotesize{\textbf{0.84}}&\footnotesize{\textbf{0.78}}&\footnotesize{\textbf{0.79}}&\footnotesize{\textbf{0.76}}&\footnotesize{\textbf{0.77}}\\
\hline \hline
\end{tabular}
\caption{Precision (P), Recall (R) and FMeasure (F1) for SVM, C4.5 and NB in the FreshMeat and SourceForge data sets.}
\end{center}
\end{table}
%\subsection{Classifiers Evaluation}
%\label{sec:classif}
\noindent \textbf{Classifiers Evaluation}. In order to measure the
classifier's performance without introducing subset selection or
feature selection bias, we have used 10-fold cross validation. The results
from the $10$ folds are averaged to produce a single
estimation. We use {\it F-measure} to estimate the quality of each
classifier. {\it F-measure} is defined as the harmonic mean
between a classifier's precision and recall. All three measures
were computed as follows:
\begin{small}
\begin{equation}
Recall = \frac{TruePos}{TruePos + FalseN},
Precision = \frac{TruePos}{TruePos + FalsePos}
\end{equation}
\begin{equation}
F-measure =  \frac{2\cdot precision \cdot recall}{precision + recall}
\end{equation}
\end{small}
where \textit{TruePos} is the number of the actual {\it
successful} projects classified as {\it successful},
\textit{FalseN} is the number of the actual {\it successful}
projects classified as {\it unsuccessful} and \textit{FalsePos} is
the number of the actual {\it unsuccessful} projects classified as
{\it successful}. Table 2 shows Precision, Recall and F-Measure
values for the 10-fold cross validation execution of the J.48,
Naive Bayes and SVM classifiers in the FreshMeat and the
SourceForge data sets respectively. SVMs managed overall the top
F-Measure compared to J.48 and the NB classifiers. For the
FreshMeat data set, the classifiers reached an F-Measure of around
$75\%$, with a precision reaching $88\%$ for the SVMs. For the
SourceForge data set, the classifiers reached an F-Measure of the
same level with an overall smaller precision from the FreshMeat
data set. In general, the classifiers performance for the
SourceForge data set is smaller than in FreshMeat, depicting that
FreshMeat's features are more descriptive for the used criterion
of porting. This is also verified from the IG and chi-square
feature values in Table 1. Overall, the proposed metric can
predict whether a project will be ported into FreeBSD and Gentoo
Portage, with high F-Measure.
\vspace{-0.2in}
\section{Conclusions and Future Work}
\vspace{-0.1in}
\label{sec:conclusions}
In this paper we propose a new Open Source Software (OSS) successfulness metric,
that is based on the development of classifiers which predict the porting of an OSS
into the FreeBSD and Gentoo Portage open source operating systems.
We have evaluated the proposed metric by measuring the performance
of Support Vector Machines (SVM), Decision Trees (C4.5) and Naive
Bayes classifiers constructed on the features contained for all
projects in FreshMeat and SourceForge. %The data used were taken from the Floss project.
We also conducted an analysis of the
features' importance and we experimentally show that the
classifiers obtain similar performance if the top-5 features are
kept, instead of all, which also include the most important
features according to previous related work. %Thus, the proposed
%metric can be based in smaller classifier models, by using
%Information Gain (IG) and chi-square as measures of feature
%selection.
As a future work we aim at combining heuristic criteria and/or
manually annotated projects to enrich the training procedure with
instances of better quality. We also aim at stacking classifiers
for the purpose of boosting the classifier's performance.

\vspace{-0.3in}
\bibliographystyle{spmpsci}
\bibliography{AISEW_2009_Quality_Classifiers}
\vspace{-1in}
\end{document}